\documentclass[10pt]{article}
\usepackage[english]{babel}
\usepackage[utf8x]{inputenc}
\usepackage[OT1]{fontenc}
\usepackage{amsfonts, amsmath, amsthm, amssymb, amsbsy}

\usepackage{multicol} 
\usepackage{stackrel}

\usepackage{graphicx}
\usepackage{listings}
\usepackage{csquotes}
\usepackage[hidelinks]{hyperref}
\usepackage[nottoc,notlot,notlof]{tocbibind}

\usepackage{subcaption}
\usepackage[margin=1in]{geometry}
\usepackage{xcolor}
\usepackage{braket}

\title{Exact Solutions for Small Systems: Urns Models}
\author{
  \parbox{\linewidth}{ {\small
    Manuel Eduardo Hernández-García\textsuperscript{1,\dag} and Jorge Velázquez-Castro\textsuperscript{1,\ddag}}\\
    {\footnotesize
    \textsuperscript{1}Facultad de Ciencias Físico Matemáticas, Benemérita Universidad Autónoma de Puebla, Heroica Puebla de Zaragoza 72570, México. \\
    \textsuperscript{\dag} \texttt{manuel.hernandezgarcia@viep.com.mx}, \textsuperscript{\ddag}\texttt{jorgevc@fcfm.buap.mx}.
  }}}
\date{\today}
\begin{document}
\maketitle  

\begin{abstract}
In this study, we analyzed urn models by solving the discrete-time master equation using an expansion in moments.  This approach is a viable alternative to conventional methods, such as system-size expansion, allowing for the determination of analytical expressions for the mean and variance in an exact form and thus valid for any system size. In particular, this approach was used to study Bernoulli-Laplace and Ehrenfest urns, for which analytic expressions describing its evolution were found.  This approach and the results will contribute to a more comprehensive understanding of stochastic systems and statistical physics for small-sized systems, where the thermodynamic limit cannot be assumed.

\end{abstract}

\tableofcontents
\section{Introduction}
Stochastic processes play a fundamental role in various fields, including biology, physics, chemistry, finance, and economics \cite{Gar, Scott, Kotz, Chandrasekhar}. The master equation is a key tool in representing and describing these processes \cite{Scott, Kotz}. It is applied to both discrete- and continuous-time stochastic processes. However, for most systems, solving their master equation is a major challenge.
In the case of continuous-time stochastic processes, the Gillespie algorithm and its variants are commonly used to simulate realizations of the process evolution \cite{Lecca}. These simulations allow for the analysis of properties, such as the mean, variance, and estimation of the probability distribution.
Urn models are common problems in discrete-time stochastic processes \cite{Scott} and are very helpful in illustrating different methods and techniques for analyzing stochastic processes. The most commonly used technique to solve this type of problem is called linear noise approximation or system-size expansion \cite{VanKampen, Jacobsen, Flajo}, which describes the fraction of balls in each urn instead of the number of balls. When the total number of balls increases, the fraction of balls in each urn can be approximated as a continuous variable, along with the time variable, leaving a partial differential equation to be solved. The solution obtained via system-size expansion is the solution in the large-size system limit. However, this approximation is not always applicable in many systems of interest, particularly in cell biology \cite{Kotz, Hayhoe, Baur, Sprott}.
Among other techniques, that do not rely on approximations are based on finding generating functions, and matrix representation \cite{Gar, Scott}, however, these methods have limitations when applied to master equations with non-polynomial transition rates.
Finally, the moment-expansion formalism \cite{Gomez, Manu}, which employs the master equation to find a set of differential equations for the moments, such as the mean and variance, without explicitly solving the master equation, has mainly been applied to continuous-time stochastic processes.  The use of moment-expansion formalism in discrete-time systems has been little exploited but is a promising technique \cite{Pickering, Sheton}.

\noindent In this study, we used the moment expansion technique to analyze the dynamics of Bernoulli-Laplace \cite{Bernoulli, Laplace} and Ehrenfest urns \cite{Ehrenfest}.
Bernoulli-Laplace urn has historical significance in probability theory \cite{Laplace} as well as the challenges and difficulties it poses in finding an analytical solution that considers discrete time \cite{Jacobsen}. This was one of the first problems related to its type. Ehrenfest urn has been applied in physics and natural sciences to understand phenomena such as particle diffusion in gases, population dispersion, and the evolution of biological systems \cite{Pickering}. For example, the Ehrenfest problem is analogous to species dispersal in an ecosystem, or particle diffusion in a porous medium \cite{Kotz}. 

\noindent We found that the moment-expansion leads to an exact solution, even for small systems. This shows that this method can be useful in finding exact solutions in mesoscopic systems and describing event-driven stochastic dynamics.  These results serve as a basis for the study of more complex systems, such as the mesoscopic and stochastic cell regulatory networks.

\noindent The remainder of this paper is organized as follows. In Section \ref{Section 2}, Bernoulli-Laplace urns are studied, and analytic expressions for the mean and variance are obtained from the master equation.  In Section \ref{Section 3}, Ehrenfest urns are studied, and analytic expressions for the mean and variance are obtained from the master equation. Section \ref{Section 4} presents the conclusions and results.

\section{Bernoulli-Laplace urns model} \label{Section 2}
The Bernoulli-Laplace urns problem was originally proposed by Bernoulli \cite{Bernoulli}. It is a stochastic process with discrete-time and discrete states space \cite{Scott}. They have applications in various fields such as physics and biology, for example, diffusion processes and gene frequencies in a finite population \cite{Kotz, Flajo, Diaconis}.

Until now, one of the most widely recognized analytical solutions involves a system size expansion, which results in a Fokker-Planck equation \cite{Jacobsen}. Other methods exist for solving similar problems; for instance, by relating differential equations to an urn model \cite{Flajo} and solving these equations directly. In this study, we employ moment expansion techniques to accurately determine the mean and variance.

\begin{figure} [h!t]
  \centering
\includegraphics[width=0.38\textwidth]{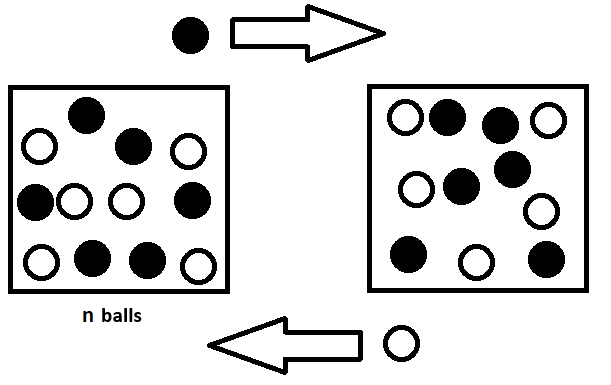}
  \caption{\textbf{Bernoulli-Laplace urns model.} In this model we have two urns, each initially containing $n$ black and $n$ white balls. A ball is taken from one urn and transferred to the other; simultaneously, the same movement is performed with the other urn. }
  \label{fig1}
\end{figure} 

The problem consists of two urns, each initially containing $n$ black and $n$ white balls. 
A ball is taken from one urn and transferred to the other; simultaneously, the same movement is performed with the other urn, as shown in Figure \ref{fig1}. We aim to determine the number of white balls in the left urn after $r$ exchanges. The master equation describes the probability $p_{x}(r)$ that there are $x$ white balls in the left urn after $r$ exchanges, and is given as follows \cite{Scott}

\begin{equation}
    p_{x}(r+1)=\left( \frac{x+1}{n} \right)^2 p_{x+1}(r) + 2\frac{x}{n} \left( 1- \frac{x}{n}\right) p_{x}(r) + \left( 1- \frac{x-1}{n}\right)^2 p_{x-1}(r). \label{1}
\end{equation}
The first term on the right side of the equation represents the probability of exchanging a white ball with a black ball. Considering that $(x+1)/n$ is the probability of choosing a white ball from the left urn when it contains $x+1$ white balls, in the right urn, there will be the same number of balls but black.  
Therefore, $(\frac{x+1}{n})^{2}$ is the probability of choosing a white ball from the left urn and a black ball from the right urn. Finally, this probability is multiplied by the probability $p_{x+1}(r)$ of having $x+1$ white balls in the right urn. Similarly, the remaining terms are constructed by noticing that $(1-x/n)$ is the probability of obtaining a white ball from the right urn.

\begin{figure}[h!t]
\begin{subfigure}{\linewidth}
\includegraphics[width=0.5\textwidth]{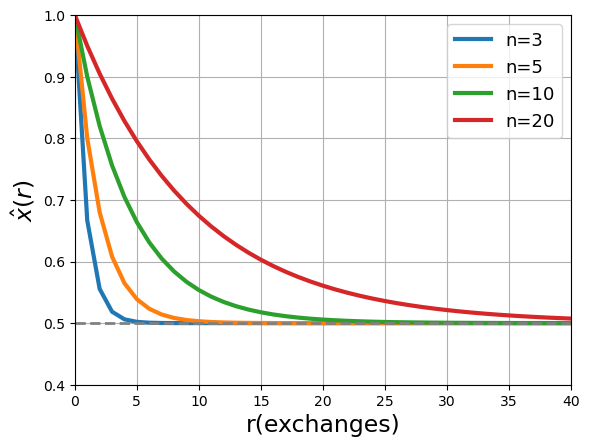}\hfill
\includegraphics[width=0.5\textwidth]{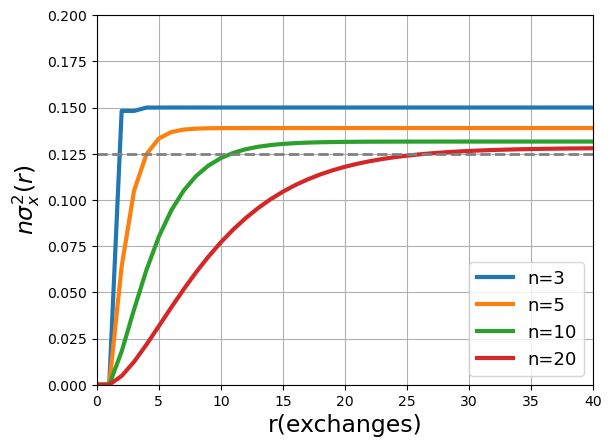}
\end{subfigure}
\caption{ \textbf{The mean and variance of Bernoulli-Laplace Urns. } In the left figure, we observe how the proportion of white balls in the first urn behaves as the iterations progress, all tending towards $\frac{1}{2}$. Meanwhile, in the right-hand figure, to compare behaviors, we multiply the variance by $n$. As $n$ increases, the values in the right-hand graph tend towards $\frac{1}{8}$.}
\label{fig2} 
\end{figure}

Multiplying the master equation (\ref{1}) by $x/n$ and taking the average, we obtain:
\begin{align}
    \hat{x}(r+1)=& \hat{x}(r) + \frac{1}{n} - \frac{2}{n} \hat{x}(r), \label{2}
\end{align}
where $\hat{x}(r)$ is the proportion of white balls in the right urn after $r$ exchanges. Considering an initial proportion $\hat{x}(0)=1$, we obtain the exact expression
\begin{align}
    \hat{x}(r)=& \frac{1}{2} \left(  1 +  \left( 1- \frac{2}{n} \right)^r \right). \label{3}
\end{align}
 For $n=1$, the system oscillates between 1 and 0. For $n=2$, $\hat{x}(r)=\frac{1}{2}$ ($r>1$), which implies that, on average, at least one white ball is in the urn. For $n>3$, $\hat{x} \to \frac{1}{2}$ as $r \to \infty$, coinciding with the steady state of the system. Figure \ref{fig2} shows a plot of the average and its tendency towards $\frac{1}{2}$. As the size of the system increases, more exchanges are required to reach equilibrium.

The variance is calculated by multiplying the master equation (\ref{1}) by $((x- \braket{x}_{r+1})/n)^2$  and summing over $x$, 

{\small
\begin{align}
    \sigma^2_x(r+1)=& \frac{1}{n^2}  - \frac{2}{n}\hat{x}(r+1) +(\hat{x}(r+1))^2 + 2\hat{x}(r) \left(\frac{1}{n} + \frac{2 \hat{x}(r+1) }{n}-2 \hat{x}(r+1)  -\frac{1}{n^2}\right) + \frac{\braket{x^2}_r}{n^2} \left(1 + \frac{2}{n^2}-\frac{4}{n}\right), \label{4}
\end{align}}
where $\sigma^2_x(r)= \frac{1}{n^2} \braket{(x- \braket{x}_{r})^2}_r$. We perform an expansion around the average for $\braket{x^2}_r$, similar to what is done in \cite{Manu},
\begin{align}
    \frac{\braket{x^2}_r}{n^2} =& \frac{1}{n^2} \braket{  \braket{x}^2_r + 2(x- \braket{x}_r) \braket{x}_r + (x- \braket{x}_r)^2  }_r \nonumber \\
                =& \hat{x}^2(r) + \sigma^2_x(r),
\end{align}
from this result and using (\ref{2}), equation (\ref{4}) simplifies to
\begin{align}
    \sigma^2_x(r+1)=& \sigma^2_x(r) + \frac{2}{n} \left( \frac{\hat{x}(r)}{n} - \frac{\hat{x}^2(r)}{n} - \sigma^2_x(r) \left( 2 - \frac{1}{n} \right) \right).
\end{align}
Assuming $\sigma^2_x(0)= 0$ and using (\ref{3}), we obtain
\begin{align}
    \sigma^2_x(r)=&   \frac{1}{8n-4} \left( 1 +  (2n-2)\left(\frac{(n-4) n+2}{n^2}\right)^r+ (1-2n)\left(\frac{n-2}{n}\right)^{2 r} \right), \label{6}
\end{align}
the variance is zero for $n=1$, whereas for $n>1$ it converges to $\frac{1}{8n-4}$, when $n$ is larger $\sigma^2_x \to \frac{1}{8n}$. This is the result of the continuous approximation for sufficiently large  $n$. This analysis can be observed in Figure \ref{fig2}, where we multiplied the variance by  $n$ to compare the variances for different  $n$. We can observe that as the value of  $n$ increases, the stationary value of the variance approaches the predicted value in the continuous case, which is plotted with a dashed line. Another aspect to consider from Equation (\ref{6}) is that, as  $n$ increases, a greater number of iterations are required to reach its steady state. Having analytical expressions for the mean and variance allows us to predict the behavior of the system, regardless of the number of exchanges ($r$) and the number of types of balls ($n$).

\section{Ehrenfest urns model} \label{Section 3}
Next, we studied the Ehrenfest urns model \cite{Ehrenfest}. Studying the Ehrenfest urn is essential in probability theory and physics, as it provides a simple yet powerful model for understanding the randomness and evolution of dynamic systems, with applications in physics and natural sciences, such as particle diffusion and population dispersion. This model teaches the key concepts of equilibrium and reversibility, develops important mathematical skills, and serves as the foundation for more complex models \cite{Redner}.

\begin{figure} [h!t]
  \centering
\includegraphics[width=0.38\textwidth]{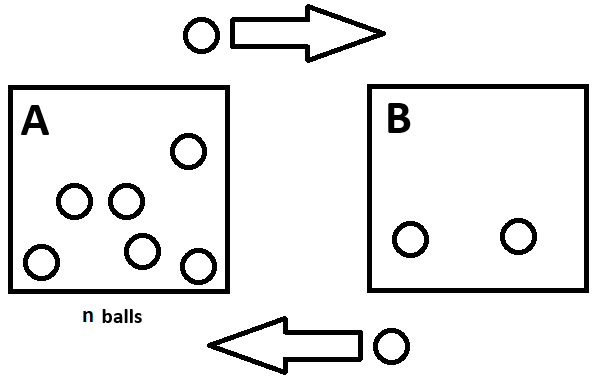}
  \caption{\textbf{Ehrenfest urns model.} This model consists of $n$ balls in one urn, whereas the other is empty, then a ball was drawn from one of the urns and transferred to the other.}
  \label{fig3}
\end{figure} 

The Ehrenfest urn model initially consists of $n$ balls in one urn, whereas the other is empty \cite{Ehrenfest}. Subsequently, a ball is drawn to be transferred to the other urn, as illustrated in Figure \ref{fig3}, where urn A initially contains $n$ balls.  The question posed by this model is how many balls will be in urn A after $r$ exchanges. The master equation describing this process is as follows:
\begin{equation}
    p_{x}(r+1)= \frac{n-x+1}{n}  p_{x-1}(r) + \frac{x+1}{n}  p_{x+1}(r), \label{7}
\end{equation}
the first term on the right-hand side describes the probability of drawing a ball from urn A and transferring it to urn B, whereas the second term describes the probability of transferring a ball from urn B to urn A.

\begin{figure}[h!t]
\begin{subfigure}{\linewidth}
\includegraphics[width=0.5\textwidth]{fig1.png}\hfill
\includegraphics[width=0.5\textwidth]{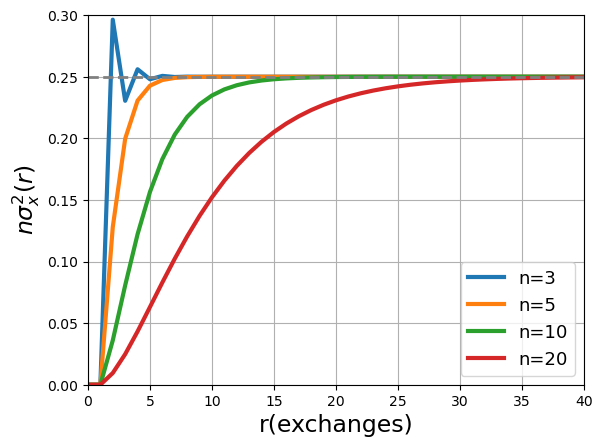}
\end{subfigure}
\caption{ \textbf{The mean and variance of Ehrenfest Urns.}  In the left figure, we observe how the proportion of white balls in the first urn behaves as the iterations progress, all tending towards $\frac{1}{2}$. Meanwhile, in the right-hand figure, to compare behaviors, we multiply the variance by $n$. As $n$ increases, the values in the right-hand graph tend towards $\frac{1}{4}$. }
\label{fig4}
\end{figure}

We follow a procedure similar to that used for Bernoulli urns. To do this, we calculate the mean and variance using the master equation (\ref{7}). First, the mean value is calculated, to do this, we multiply the master equation by $\frac{x}{n}$, then we sum over $x$ and consider the initial proportion $\hat{x}(0)=1$, thus we obtain an exact expression
\begin{align}
    \hat{x}(r)=& \frac{1}{2} \left(  1 + \left( 1- \frac{2}{n} \right)^r \right), \label{9}
\end{align}
where $\hat{x}(r)$ ($=\frac{1}{n}\braket{x}_r$) is the proportion of white balls in urn A after $r$ exchanges. For $n=1$, the system oscillates between 1 and 0; for $n=2$, $\hat{x}(r)=\frac{1}{2}$ ($r>1$), implying that on average, at least one ball is in one of the urns. For $n>3$, $\hat{x} \to \frac{1}{2}$ as $r \to \infty$, coinciding with the steady state of the system. In Figure \ref{fig4}, we can observe how the average is plotted and how it tends to $\frac{1}{2}$ as the system becomes larger, thus requiring more exchanges.

Next, we determine the variance. To do this, we used the master equation (\ref{7}). Following a process similar to that described in the previous section, and assuming $\sigma^2_x(0)= 0$, we obtain:
\begin{align}
    \sigma^2_x(r)=& \frac{1}{4n}  \left(1 + (n-1) \left(1 -\frac{4}{n}\right)^r -n \left(1 -\frac{2}{n}\right)^{2 r}\right), \label{11}
\end{align}
where $\sigma^2_x(r)= \frac{1}{n^2} \braket{(x- \braket{x}_{r})^2}_r$. For $n=1$, the variance is zero. For $n=2$, the variance oscillates between zero and $\frac{1}{4}$, whereas for $n>3$, it converges to $\frac{1}{4n}$. Figure \ref{fig4} illustrates the variance behavior. To compare the variances for different $n$, we multiply the variance by $n$. We can observe that, as the value of $n$ increases, the system requires more iterations to reach its steady state.

We observe that the Ehrenfest urn model has the same mean as the Bernoulli urn model, which essentially means that in both models, the number of white balls in the urns becomes uniform. However, the way the balls are exchanged and the fact that in the Bernoulli-Laplace urn model, there are actually black balls make the difference evident.

\section{Conclusions and Results} \label{Section 4}

The application of moment expansion results in difference equations describing the exact evolution of the mean and variance. This approach is valuable for stochastic systems with discrete time, because it enables a detailed analysis of the system's behavior in terms of its moments without resorting to any continuous-time approximation. This underscores the significance of our findings in providing a practical and insightful approach for analyzing and predicting the behavior of complex systems in the mesoscopic regime. Specific for both Bernoulli-Laplace and Ehrenfest urns, we found analytic exact expressions for the mean and variance for any system size, that coincide with those of continuous approximation when the number of balls is larger and a continuous time approximation \cite{Scott}.  In future work, extensions of this study could be explored, such as the inclusion of nonlinear interactions or consideration of higher-order moments. Nonlinear interactions are common in many systems and their incorporation can provide a more accurate representation of real-world phenomena. Additionally, investigating higher-order moments can offer deeper insights into a system's behavior, capturing more nuanced aspects of variability and dynamics. Such extensions will contribute to a more comprehensive understanding of stochastic systems and statistical physics for small-sized systems, where the thermodynamic limit cannot be assumed.

\section*{Acknowledgments}
Manuel E. Hernández-García acknowledges the financial support of CONAHCYT through the program "Becas Nacionales 2023".\\
Jorge Velázquez-Castro acknowledges Benemérita Universidad Autónoma de Puebla-VIEP financial support through project 00398-PV/2024.

\section*{Declarations}
The authors declare no conflicts of interest regarding the publication of this article. 

\noindent All data generated or analyzed in this study are included in this published article.

\end{document}